\documentclass[12pt]{article}

\usepackage{geometry}
\usepackage{microtype}
\usepackage{amsmath,amsthm,amsfonts,amssymb}
\usepackage{bbold}
\usepackage[mathscr]{eucal}
\usepackage{xcolor}
\usepackage{mathtools}

\usepackage{fouriernc}

\newtheorem{thm}{Theorem}
\newtheorem{lmm}{Lemma}

\title{Hypergeometric First Integrals of the Duffing and van der Pol Oscillators}

\author{Tomasz Stachowiak\footnote{tomasz@amp.i.kyoto-u.ac.jp}\\
\small Department of Applied Mathematics and Physics,\\
\small Graduate School of Informatics, Kyoto University,\\
\small 606-8501 Kyoto, Japan}
\date{}

\begin{document}

\maketitle

\begin{abstract}
The autonomous Duffing oscillator, and its van der Pol modification, are known to admit time-dependent first integrals for specific values of parameters. This corresponds to the existence of Darboux polynomials, and in fact more can be shown: that there exist Liouvillian first integrals which do not depend on time. They can be expressed in terms of the Gauss and Kummer hypergeometric functions, and are neither analytic, algebraic nor meromorphic. A criterion for this to happen in a general dynamical system is formulated as well.
\end{abstract}

\section{}

The Duffing and the van der Pol oscillators are among the simplest (at least in form) dynamical systems, present in biology \cite{FitzHugh,Nagumo}, nano-electronics and mechanics \cite{Farid} as well as many subfields of physics \cite{Brennan,Rajaseekar} (see the last one for a more comprehensive list of references). The systems' integrability, or solvability, is still an active topic with only partial results available \cite{Parth,Chandra,Gao} and the aim of the present article is to use them as examples of how non-analytic Liouvillian first integrals can arise in polynomial differential equations. This is the content of Lemma 1 in Section 2 and its generalization in Section 5.

The force-free form of the basic Duffing oscillator is
\begin{equation}
    \ddot{u} + \dot{u} +\omega_0^2 u + u^3 = 0,
    \label{main_duff}
\end{equation}
or, more generally,
\begin{equation}
    \ddot{u} + \alpha\dot{u} +\omega_0^2 u + \gamma u^n = 0.
    \label{gen_duff}
\end{equation}
With a suitable rescaling of $u$ and the time, $\alpha$ and $\gamma$ can be made equal to 1, so the only essential parameters will be $\omega_0$ and $n$. Although the case of immediate interest is that of integer $n>1$, the extension to real $n$ is straightforward if $u^n$ is replaced by $|u|^{n-1}u$. 

For the specific value of $\omega_0^2=\tfrac29$, explicit solutions of \eqref{main_duff} in terms of Jacobian elliptic functions were found in \cite{Parth}. The authors note that the equation admits a transformation
\begin{equation}
    W = \frac{3}{\sqrt{2}}\mathrm{e}^{t/3} u;\quad
    Z = -\sqrt{2}\mathrm{e}^{-t/3},
    \label{chandra_trans}
\end{equation}
which turns the equation into a solvable one
\begin{equation}
    W''(Z) = - W^3.
\end{equation}
It can be regarded as Hamiltonian with
\begin{equation}
    H = \tfrac12 W'^2 + \tfrac14 W^4, \label{hamil}
\end{equation}
which is then conserved, but upon going back to $u$ and $\dot{u}$, the first integral defined by $H$ becomes time dependent:
\begin{equation}
    I_{\mathrm{D}} = \mathrm{e}^{\frac43 t}\left(\frac12 \dot{u}^2+\frac13u\dot{u}
    +\frac{1}{18}u^2+\frac14 u^4\right).
    \label{t_int}
\end{equation}
A generalization of the above approach, which utilizes transformations of the form \eqref{chandra_trans}, was formulated in \cite{Chandra}, and gives a more constructive method of finding first integrals.

As the solutions are explicitly available, this time dependence of $I_{\mathrm{D}}$ does not seem to be a practical problem -- one could argue that a first integral can be obtained (at least locally) by elimination of time between the solutions, but that is seldom practically computable. Thus, the question emerges of whether a ``proper'' time-independent first integral exists. The goal of this investigation is to show that this is indeed the case for the general unforced Duffing oscillator, and partly so for the Duffing-van der Pol oscillator. What is more, the existence of such an integral can be ascertained without the {\it a priori} knowledge of the solutions.

The first main result of the present letter is the following 
\begin{thm}
If the frequency $\omega_0$ of the Duffing oscillator \eqref{gen_duff} with general nonlinear term of degree $n$ satisfies
\begin{equation}
    \omega_0^2 = \frac{2(n+1)}{(n+3)^2},
\end{equation}
then the equation has a Liouvillian first integral. When $n$ is odd, in the half-planes separated by $2u+(3+n)\dot{u}=0$ the integral is given by
\begin{equation}
    I_1 = V^{-\frac{1}{n+1}}\left[V^{\frac12}+(n-1)\omega_0\varsigma u\,
    _2F_1\left(\frac{1}{2},\frac{1}{n+1},\frac{n+2}{n+1},
    \frac{4u^{n+1}}{V}\right)\right],
    \label{thm1_f1}
\end{equation}
where $V=4u^{n+1}+(2u+\omega_0^2(n+3)\dot{u})^2$ and $\varsigma=\mathrm{sign}(2u+(n+3)\dot{u})$; alternatively, in the half-planes separated by $u=0$ it is
\begin{equation}
    I_2 = V^{\frac12-\frac{1}{n+1}}-
    \frac{n-1}{n+1}2^{\frac{n-1}{n+1}}\omega_0^2\varsigma' 
    \frac{2u+(n+3)\dot{u}}{\sqrt{V}}\,
    _2F_1\left(\frac{1}{2},\frac{n}{n+1},\frac32,1-\frac{4u^{n+1}}{V}\right),
    \label{thm1_f2}
\end{equation}
with $\varsigma' =\mathrm{sign}(u)$ and $V$ as above. Each ${}_2F_1$ is real when its argument is less than 1.

When $n$ is even, so that $V=0$ defines a real invariant curve $\mathcal{C}$, the above expressions remain valid if $I_1$ is taken to the right of $\mathcal{C}$, and $I_2$ to its left with $\varsigma'=(-1)^{n/(n+1)}$.
\end{thm}

The second equation of interest is the Duffing-van der Pol oscillator, which exhibits the same type of time-dependent first integrals. The equation is
\begin{equation}
    \ddot{u} + (1+\beta u^m)\dot{u}+\omega_0^2 u + u^n = 0,
    \label{vanderpol_gen}
\end{equation}
and it was shown in \cite{Gao} to admit a whole parametric family of time-dependent integrals. The required condition is that $m=n-1$ and $n^2+\beta^2\omega_0^2 = n\beta$, and the conserved quantity is
\begin{equation}
    I_{\mathrm{P}} = \mathrm{e}^{\frac{n}{\beta}t}\left(\dot{u}+\frac{\beta-n}{\beta}u
    +\frac{\beta}{n}u^n\right).
    \label{ipe}
\end{equation}
Similarly to the previous system, this quantity can be used to obtain a time independent first integral, leading to the second result:
\begin{thm}
If the parameters of the Duffing-van der Pol oscillator \eqref{vanderpol_gen} with general nonlinear term of degree $n$ satisfy
\begin{equation}
    \omega_0^2 = \frac{n}{(n+1)^2},\quad
    \beta = n+1 \quad\text{and}\quad m=n-1,
    \label{th2_cond}
\end{equation}
then the equation has a Liouvillian first integral given by
\begin{equation}
    I_3 = \left|u^n+\omega_0^2(u+\beta \dot{u})\right|^{1-\frac{1}{n}}
    \left[1+\frac{(n-1)u}{u+\beta\dot{u}}\, 
    _2F_1\left(1,1,1+\frac{1}{n},
    -\frac{u^n/\omega_0^2}{u+\beta\dot{u}}\right)\right],
    \label{th2_f1}
\end{equation}
which is real when $-\omega_0^{-2}u^n/(u+\beta\dot{u})<1$;
or alternatively by
\begin{equation}
    I_4 = \left|u^n+\omega_0^2(u+\beta \dot{u})\right|^{1-\frac{1}{n}}
    \left[1-\frac{u}{u+\beta\dot{u}}\, 
    _2F_1\left(1,1,2-\frac{1}{n},
    1+\frac{u^n/\omega_0^2}{u+\beta\dot{u}}\right)\right],
    \label{th2_f2}
\end{equation}
which is real when $\omega_0^{-2}u^n/(u+\beta\dot{u})<0$. Additionally, the singularities of the hypergeometric function correspond to the invariant sets $\omega_0^2(u+\beta\dot{u})+u^n=0$ and $u+\beta\dot{u}=0$.
\end{thm}

The above system can be further generalized to
\begin{equation}
    \ddot{u} + (1+\beta u^{n-1})\dot{u}+\omega_0^2 u+
    u^n+\varphi u^{2n-1} = 0,\label{gen_pol}
\end{equation}
which was analyzed with the Prelle-Singer procedure in \cite{Chandra2}, and with the Lie symmetry method in \cite{Feng}. It differs from the first two in that the specific structure of its invariant sets makes the hypergeometric first integral confluent:
\begin{thm}
If the parameters of the generalized Duffing-van der Pol oscillator \eqref{gen_pol} satisfy
\begin{equation}
    \omega_0^2 = \frac{n}{(n+1)^2},\quad \beta=n+1,\quad
    \text{and}\quad \varphi = \frac{1}{4\omega_0^2},
\end{equation}
then the equation has a Liouvillian first integral given by
\begin{equation}
    I_5 = |V|^{-\frac{1}{n}}\left[\exp\left(\frac{2(n-1)\varphi u^n}{nV}\right)V+
    (n-1)u{}_1F_1\left(\frac{1}{n},1+\frac{1}{n},
    \frac{2(n-1)\varphi u^n}{nV}\right)\right],
\end{equation}
where $V=u+\beta\dot{u}+2\varphi u^n$. The above integral is real, and its only singularity corresponds to the invariant curve $V=0$.
\end{thm}

The appearance of hypergeometric functions is quite remarkable -- it shows how restrictive the notion of analytic, meromorphic or algebraic integrability can be. Although various parametric forms of solutions of polynomial dynamical systems can be given in terms of hypergeometric functions \cite{Maier}, this seems to be the first example of a natural (i.e. not artificially contrived) system with first integrals of such form. This is the price we have to pay for the transition from relatively simple, but time-dependent, integrals to time-independent ones, which permit us to call an autonomous system integrable in the standard sense.

A starting point in integrability analysis might be looking in the class of analytic functions, but because trajectories converge to the node at the origin, this approach will not work here. As the systems are of dimension two, it is not possible to directly apply the basic methods of differential Galois theory \cite{Morales} either, because the first normal variational equation will be of order 1 and thus soluble. Higher variational equations could be used, but in addition to complicating the analysis, it will still be limited to meromorphic first integrals. To make progress, we will have to turn to the Darboux polynomials, which are most often used in the context of rational first integrals, but can lead to algebraic or even Liouvillian expressions.

The remainder of the article is devoted to introducing the basic machinery (Section 2.) and providing a general criterion for the appearance of hypergeometric first integrals in polynomial systems (Lemma 1). The proofs of the theorems are given in Sections 2 through 4, and the central criterion is further extended in Lemma 2. The possibility of reduction to more elementary functions is discussed in the final section.

\section{}

The first step in the construction of first integrals is to consider the dynamical system corresponding to \eqref{main_duff}
\begin{equation}
\begin{aligned}
    \dot{x} &= x y &=:P,\\
    \dot{y} &= -\omega_0^2 - y(1+y) - x^2 &=:Q,
    \label{sys_duff}
\end{aligned}
\end{equation}
where $x=u$ and $ y=\dot{u}/u$. The vector field associated to the flow defines a derivation over the polynomial ring $\mathbb{C}[x,y]$, and one can look for the Darboux polynomials of the derivation $D:=P\partial_x+Q\partial_y$. If enough of them are found, the second step is to combine them into a first integral.

To briefly review the general setting, consider a $d$-dimensional system or, equivalently, a derivation over complex polynomials of $d$ variables. A {\it Darboux polynomial} $F$ is defined as an elements of $\mathbb{C}[x_1,\ldots,x_d]$, such that
\begin{equation}
    DF = K F,\quad\text{for some }K\in\mathbb{C}[x_1,\ldots,x_d],
\end{equation}
and $K$ is called the {\it cofactor}. Since along a solution one has $DF=\dot{F}$, it follows from the definition that $F=0$ is an invariant set, and if $K\equiv0$ then $F$ is just a polynomial first integral. Another basic property holds for the product: $D(F_1F_2)=(K_1+K_2)F_1F_2$, so that $F_1 F_2$ is itself a Darboux polynomial. In particular, any integral power $F^m$ is a Darboux polynomial with the cofactor $mK$. A converse result can also be proven: if $F$ is reducible, its factors must be Darboux polynomials too. 

Importantly, the existence of such polynomials allows one to analyze integrability through rational functions, because if $f/g\in\mathbb{C}(x_1,\ldots,x_d)$ is a first integral, then necessarily $f$ and $g$ must be Darboux polynomials with the same cofactor; conversely, if there are enough Darboux polynomials, so that $K$'s are linearly dependent over $\mathbb{Z}$, a rational first integral exists. For the full Darboux theorem see \cite{Singer}.

Going beyond polynomials, $F^{q}$, with rational $q$, is an algebraic function which still has a polynomial cofactor. Indeed for any function $f$, Liouvillian over $\mathbb{C}(x_1,\ldots,x_d)$, there might exist a polynomial $K$ such that $Df = Kf$; for brevity the term {\it Darboux pair} will be used to refer to $\{f,K\}$ then.

The last general remark to be made is that the degree of $K$ has to be lower than that of the derivation, and finding variables for which $D$ is quadratic will simplify things considerably. However, neither this, nor $d=2$ is assumed for the general lemmas given below. For an exposition of the subject in the context of polynomial derivations see also \cite{Nowicki}.

Coming back to the present case of \eqref{sys_duff}, the derivation is of degree 2, so the cofactors can be at most linear, and assuming some degree of $F$, the equation $DF=KF$ can be solved term by term. It is a straightforward calculation to verify that when $\omega_0^2=\tfrac29$ the following quadratic Darboux polynomials exist
\begin{equation}
\begin{aligned}
    F_1 &= x, &K_1 &= y,\\
    F_2 &= 9x^2+2(1+3y)^2, &K_2 &= -\tfrac43 -2y.
    \label{dars1}
\end{aligned}
\end{equation}
As mentioned, the irreducible factors of $F_2$ are themselves Darboux polynomials
\begin{equation}
    F_{3,4} = 2\pm3\sqrt{2}\mathrm{i} x+6y,\quad
    K_{3,4} = -\tfrac23 \pm\sqrt{2}\mathrm{i}x-y,
\end{equation}
but with a view to finding a real first integral, $F_2$ will be used.

It should also be noted, that there is no effective general tool to give the bounds on the degree of $F$, so when $\omega_0^2\neq\tfrac29$ the question of existence of higher degree $F$ remains open. In general it is a difficult task to exclude all possible degrees, see for example \cite{Maciejewski}, but fortunately the goal here is not to find all the polynomials or rational first integrals.

Thanks to the product property, Darboux polynomials \eqref{dars1} can be combined, and the new cofactor can be made constant
\begin{equation}
\frac{D\left(x^2(9x^2+2(1+3y)^2)\right)}{x^2(9x^2+2(1+3y)^2)} 
= -\frac43,
\end{equation}
so that we immediately have
\begin{equation}
    J := x^2(9x^2+2(1+3y)^2) = \mathrm{e}^{-4t/3}J_0,
\end{equation}
where $J_0=J(x(0),y(0))$ is equivalent to $I_{\mathrm{D}}$ of \eqref{t_int}.

If $K$'s are dependent over $\mathbb{Z}$, the resulting cofactor can even be made zero, so a rational first integral is found (or at least algebraic, were they dependent over $\mathbb{R}$). This is not the case here, and there are no new Darboux polynomials of degree 3 either, that is, they are all products of $F_1$ and $F_2$. In principle, one could try looking at higher degrees, but as it turns out this will not be necessary. 

An important thing to notice is also that the following combination of \eqref{dars1} is related to the divergence of the flow \eqref{sys_duff}
\begin{equation}
    D(F_1/F_2)= -(\partial_xP+\partial_yQ)(F_1/F_2).
\end{equation}
This means that a Liouvillian first integral can be obtained, as described by Theorem 1 of \cite{Singer} with $F_1/F_2$ being the integrating factor of the form 
$P\mathrm{d}y-Q\mathrm{d}x$. Instead of applying that theorem, however, a slightly different derivation will be given here, which leads directly to a more concise expression for the first integral. It relies on a result valid in a very general setting:

\begin{lmm} Let $D=\sum_{i=1}^{d} P_i\partial_i$ be a derivation which admits two Darboux polynomials $f_1$ and $f_2$ with cofactors $k_1$ and $k_2$, respectively. 
If $f_3:=f_2-f_1$ is also a Darboux polynomial with cofactor $k_3$, such that
\begin{equation}
    \alpha_1 k_1 +\alpha_2 k_2 +\alpha_3 k_3=\alpha_0\in\mathbb{C},\qquad
    \alpha_i\in\mathbb{C}\;\;\text{not all zero},
\end{equation}
and $f_3$ is the cofactor of $f_1/f_2$, i.e.,
\begin{equation}
    k_1-k_2 = f_3,
\end{equation}
then the dynamical system $\dot{x}=P_i(x)$ associated with the derivation $D$ has a Liouvillian first integral, expressible locally in terms of the hypergeometric function.
\end{lmm}
{\it Proof.} The first condition means that we have a function
$J=\prod f_i^{\alpha_i}$ such that $DJ=\alpha_0 J$ or 
$J(t)=\mathrm{e}^{\alpha_0 t}J(0)$. Let us thus take the time integral of $J$ and re-express it with $f_i$ and a new variable $\zeta := f_1/f_2$ to get
\begin{equation}
\begin{aligned}
    \int J\mathrm{d}t &= 
    \int f_1^{\alpha_1}f_2^{\alpha_2}f_3^{\alpha_3}\mathrm{d}t,\\
    \frac{J}{\alpha_0} + C_1 &=
    \int f_1^{\alpha_1}f_2^{\alpha_2}f_3^{\alpha_3}
    \frac{\mathrm{d}\zeta}{( f_2- f_1)\zeta}=
    \int (f_1/f_2)^{\alpha_1}f_2^{\alpha_1+\alpha_2-1}f_3^{\alpha_3}
    \frac{\mathrm{d}\zeta}{(1-\zeta)\zeta},\\
    \frac{J}{\alpha_0} + C_1 &=
    \int \zeta^{\alpha_1-1}(1-\zeta)^{\alpha_3-1}
    f_2^{\alpha_1+\alpha_2+\alpha_3-1}\mathrm{d}\zeta,
\end{aligned}
\end{equation}
where the determination of the complex argument of $\zeta$ or $1-\zeta$ was ignored, with the understanding that the form of the integral remains the same save for a multiplicative constant. This constant depends on the region in which $\zeta$ (and hence $x$) lies, which is why the word ``locally'' is necessary in the conclusion.

Because $\alpha_i$ are defined up to rescaling, we can take $\alpha_1+\alpha_2+\alpha_3=1$ and the last integral becomes the Euler representation of the hypergeometric function, so that
\begin{equation}
    I = \frac{J}{\alpha_0} - 
    \frac{\zeta^{\alpha_1}}{\alpha_1}\,
    _2F_1(1-\alpha_3,\alpha_1,\alpha_1+1,\zeta),
\end{equation}
is the sought first integral. Because the Euler representation is obtained from an indefinite integral, the result is a Liouvillian function of $\zeta$, which in turn is a rational function of the original variables. $\qed{}$
 
Thus, in the standard case of equation \eqref{main_duff}, the above lemma can be applied with
\begin{equation}
    f_1 = -\tfrac13 F_4,\quad f_2 = -\tfrac13 F_3,\quad 
    f_3 = 2\sqrt{2}\mathrm{i}F_1,
\end{equation}
but the resulting first integral will then explicitly contain the imaginary unit. A slight modification is required if one insists on real expressions, and it can be effected for the more general equation \eqref{gen_duff}.

{\it Proof of Theorem 1.} Let us introduce a new set of variables $x=u^{n-1}$ and $y=\dot{u}/u$, in which system \eqref{gen_duff} reads
\begin{equation}
\begin{aligned}
    \dot{x} &= (n-1)y x,\\
    \dot{y} &= -\omega_0^2 -y(1+y) - x.
\end{aligned}
\end{equation}
As before, $F_1=x$ is a Darboux polynomial, and direct computation shows that
\begin{equation}
    F_2 = 4x + \omega_0^2(2+(n+3)y)^2
\end{equation}
is another one if the condition $\omega_0^2=2(n+1)/(n+3)^2$ is satisfied. As above, a function $J$ with constant logarithmic derivative can be found to be
\begin{equation}
    J = F_1^{\frac{2}{n-1}}F_2,\quad \text{with}\quad
    D J = -(n+3)\omega_0^2 J.
\end{equation}
The new variable to use is
\begin{equation}
    \zeta = \frac{4x}{4x+\omega_0^2(2+(n+3)y)^2},
\end{equation}
for which we have
\begin{equation}
    \dot{\zeta} = D\zeta = \frac{n+1}{n+3}(2+(n+3)y)\zeta,
\end{equation}
and additionally
\begin{equation}
    \omega_0^2(2+(n+3)y)^2 = F_2 - F_1 = F_2(1-\zeta).
\end{equation}
This makes it possible to calculate the time integral of $J^{\alpha}$, with $\alpha$ to be determined,
\begin{equation}
\begin{aligned}
    \int J^{\alpha}\mathrm{d}t &= 
    \int F_1^{\frac{2\alpha}{n-1}}F_2^{\alpha}\,\mathrm{d}t
    =\int \left(\frac{F_1}{F_2}\right)^{\frac{2\alpha}{n-1}}
    F_2^{\alpha\gamma} \frac{\mathrm{d}\zeta}{\dot{\zeta}},\\
    C_1-\frac{J^{\alpha}}{(n+3)\omega_0^2\alpha} &=
    \varsigma\frac{n+3}{n+1}4^{\frac{2\alpha}{1-n}}\omega_0^{2\alpha\gamma}
    \int \zeta^{\frac{2\alpha}{n-1}-1}(1-\zeta)^{-\alpha\gamma}
    (2+(n+3)y)^{2\alpha\gamma-1}\mathrm{d}\zeta,
\end{aligned}
\end{equation}
where $\gamma = (n+1)/(n-1)$, and $|\varsigma|=1$ is a sign factor required because in this formal manipulation of exponents we have ignored the usual problems arising from complex arguments. It will be more convenient to determine it later, in the final variables $u$ and $\dot{u}$.

Taking $\alpha = 1/(2\gamma)$ in the above leads again to the Euler integral and the hypergeometric functions
\begin{equation}
\begin{aligned}
    C_1 &= \frac{n+3}{n-1}J^{\alpha} -\varsigma\frac{n+3}{n+1}2^{\frac{n-1}{n+1}}\omega_0(1-\zeta)^{\frac12}\,
    _2F_1\left(\frac12,\frac{n}{n+1},\frac32,1-\zeta\right),\quad 
    \text{or}\\
    \widetilde{C}_1 &= \frac{n+3}{n-1}J^{\alpha} +\varsigma(n+3)\,
    4^{-\frac{1}{n+1}}\omega_0\zeta^{\frac{1}{n+1}}\,
    _2F_1\left(\frac12,\frac{1}{n+1},\frac{n+2}{n+1},\zeta\right).
\end{aligned}
\end{equation}
These are two forms of the same first integral, but the former is more convenient around $\zeta = 1$, which corresponds to the line $2+(n+3)y=0$, and has a singularity at $\zeta=0$; conversely, the latter works around $\zeta = 0$ ($x=0$) and has a singularity at $\zeta = 1$.

When re-expressing $z$ and $J$ in terms of $x$ and $y$ and then $u$ and $\dot{u}$, we again ignore the complex arguments and expand the powers freely. For $\widetilde{C}_1$ ($C_1$ is analogous) this gives
\begin{equation}
    I_1 = \frac{n+3}{n-1}V^{-\frac{1}{n+1}}\left[V^{\frac12}+
    \varsigma(n-1)\omega_0 u\,
    _2F_1\left(\frac{1}{2},\frac{1}{n+1},\frac{n+2}{n+1},
    \frac{4u^{n+1}}{V}\right)\right],
    \label{proof1_f1}
\end{equation}
with $V=4u^{n+1}+\omega_0^2(2u+(n+3)\dot{u})^2$.
Instead of dealing with all the regions of $\zeta$ or $x$ and $y$ to determine $\varsigma$ it seems easier to differentiate $I_1$ and ensure that the pseudo-polynomial expression in $u$ and $\dot{u}$ vanishes identically. This happens for
\begin{equation}
    \varsigma = \frac{2u+(n+3)\dot{u}}{\sqrt{V}}
    \left(\frac{(2u+(n+3)\dot{u})^2}{V}\right)^{-1/2}.
\end{equation}
When $n$ is odd, $V$ is always positive, so this reduces to $\varsigma=\text{sign}(2u+(n+3)\dot{u})$; when $n$ is even, $V=0$ defines a real invariant curve $\mathcal{C}$, which lies in the left half-plane and has a cusp at the origin. $V$ is positive to the right of this curve, so \eqref{proof1_f1} applies there with the same sign factor. For $C_1$, and consequently $I_2$ in \eqref{thm1_f2}, the factor becomes
\begin{equation}
    \varsigma' = u \left(\frac{u^{n+1}}{V}\right)^{-\frac{1}{n+1}}
    V^{-\frac{1}{n+1}},
\end{equation}
which is just $\text{sign}(u)$ for odd $n$; when $n$ is even, $I_2$ is to be used to the left of $\mathcal{C}$, where both $u$ and $V$ are negative, so $\varsigma'$ simplifies to $(-1)^{n/(n+1)}$ as in the theorem. Both hypergeometric functions are real when their argument is less than 1, so only in the last case does the integral acquire a (constant) complex phase via the roots of $V$.\qed{}

An interesting difference between between this and Lemma 1 is that we have only used a relation between $K_1-K_2$ and $F_2-F_1$, without assuming that the latter is linked with $F_3$. Because $J$ involved only $F_1$ and $F_2$ that was enough to express the integrand as a function of $\zeta$ only. This observation will be crucial in proving a more general lemma in section 4.

It should also be noticed that the Gaussian integral happens to be expressible via the incomplete beta function, namely
\begin{equation}
    \int_0^z \zeta^{\frac{1}{n+1}-1}(1-\zeta)^{-\frac12}\mathrm{d}\zeta = 
    B_z\left(\tfrac{1}{n+1},\tfrac12\right).
\end{equation}
Together with the special case $n=3$ discussed in the introduction, this suggests a link with inverse elliptic functions or even reduction to elementary functions, which is discussed for all three systems in Section 5. 

\begin{figure}[ht]\centering
\includegraphics[width=.45\textwidth]{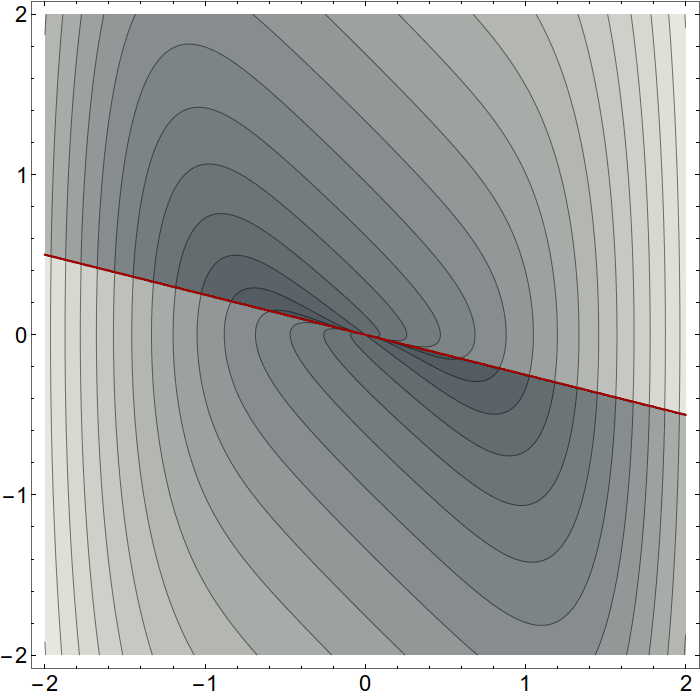}\hspace{7mm}
\includegraphics[width=.45\textwidth]{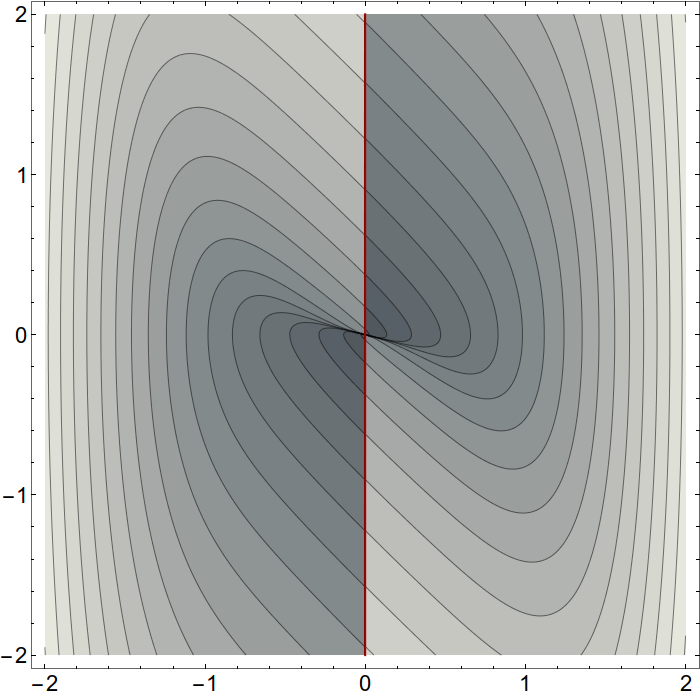}
\caption{\small The phase space of the Duffing oscillator with $n=5$ in the $u$ and $\dot{u}$ coordinates. The left diagram shows level sets of $I_1$, the right one those of $I_2$, as defined in Theorem 1. Notice the mismatch between gradient directions.}
\label{fig1}
\end{figure}

\begin{figure}[ht]\centering
\includegraphics[width=.45\textwidth]{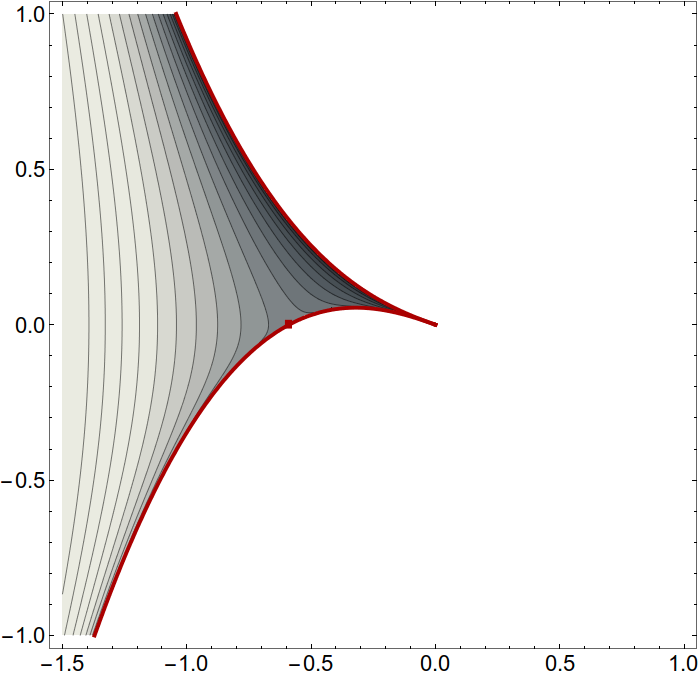}\hspace{7mm}
\includegraphics[width=.45\textwidth]{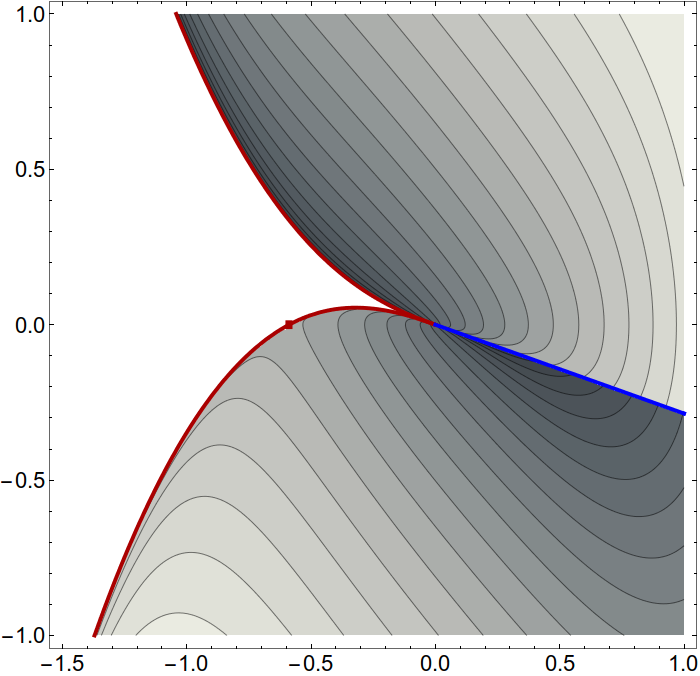}
\caption{\small The phase space of the Duffing oscillator with $n=4$ in the $u$ and $\dot{u}$ coordinates; $I_1$ and $I_2$ of Theorem 1 are used in the right and left diagrams, respectively. The discontinuity line is drawn in blue to distinguish it from the invariant curve in red. The additional critical point is depicted by a red square.}
\label{fig1a}
\end{figure}

An example of the phase portrait as expressed via levels of the first integrals for odd $n$ is shown in Figure \ref{fig1}. As the origin is an attracting node, there cannot exist a global analytic first integral. Upon passing between the half-planes, the values of the two integrals $I_1$ and $I_2$ need to be adjusted if one wants to keep the formulae intact, alternatively they can be regarded as multivalued functions. The slightly more complicated situation with even $n$ is shown in Figure \ref{fig1a}, in addition to the line of discontinuity there appears an invariant curve with another critical point on it.

Finally, one has to remember that to generalize the $u^3$ term in \eqref{main_duff} to arbitrary exponents, while keeping the ``harmonic'' character of the force, one should take $|u|^{n-1} u$ instead of just $u^n$. The argument of the hypergeometric function, which corresponds to $\zeta$, then satisfies $0\leq 4|u|^{n+1}/V\leq1$ everywhere except the origin, and the situation is again like that in Figure \ref{fig1}.

\section{}

All of the above can be immediately applied also to the Duffing-van der Pol equation \eqref{vanderpol_gen}, although just the existence of a time-dependent first integral will not be enough. As in \cite{Gao}, the exponents will have to satisfy $m=n-1$, so that excludes the classical van der Pol ($m=2$, $n=1$), and Duffing ($m=0$, $n=3$) oscillators. 

{\it Proof of Theorem 2.} Adopting the same variables as before, i.e, $x=u^{n-1}$ and $y=\dot{u}/u$, the system is now
\begin{equation}
\begin{aligned}
    \dot{x} & = (n-1)yx,\\
    \dot{y} &= -\omega_0^2 -y(1+y) -x(1+\beta y).
\end{aligned}
\end{equation}
The Darboux polynomial $F_1=x$ is self-evident, and if the condition
\begin{equation}
    \omega_0^2 = n(\beta-n)\beta^{-2},
    \label{cond1}
\end{equation}
holds, a second linear one can be found by direct computation:
\begin{equation}
    F_2 = \beta^2 x +n\beta(1+y)-n^2,\quad K_2 = -(n+\beta y)\beta^{-1}.
\end{equation}
These two are enough for the construction of \eqref{ipe}, because we have
\begin{equation}
    \frac{D(xF_2^{n-1})}{xF_2^{n-1}} = \frac{n(1-n)}{n+1}.
\end{equation}

However, this alone is not enough to use Lemma 1 because the cofactor of $F_1/F_2$, or a similar combination, is a function of $y$ alone so not necessarily a Darboux polynomial. At the same time, $F_2 -\beta^2 F_1$, which could be the candidate for $f_2-f_1$, is linear in $y$ but in order for it to be a Darboux polynomial, an additional condition is necessary:
\begin{equation}
    \omega_0^2 = (\beta -1)\beta^{-2}.
    \label{cond2}
\end{equation}
It is independent of the one in \eqref{cond1}, and it guarantees the existence of the Darboux pair
\begin{equation}
    F_3 = 1+\beta y,\quad K_3 = \beta^{-1}-1-y-\beta x.
\end{equation}
Notice that just $F_1$ and $F_3$ are insufficient for the time-dependent integral under consideration, because no linear combination of their cofactors can be made constant. It should also be said that unless a full characterization of the system's Darboux polynomials is given, it remains an open question if the condition \eqref{cond1} alone is not enough to proceed with the proof.

If both \eqref{cond1} and \eqref{cond2} are to be satisfied, it follows that $\beta=n+1$ and the conditions \eqref{th2_cond} are obtained. The situation thus resembles that of the Duffing case because the value of $\omega_0$ is strictly determined by the degree $n$.

Is is now straightforward to take
\begin{equation}
    f_1 = -(n+1)F_1,\quad f_2 = n F_3/(n+1),\quad f_3 = F_2/(n+1),
\end{equation}
and check that Lemma 1 can be applied with $\zeta = f_1/f_2$ and
$J=f_1^{1/n}f_3^{1-1/n}$, which leads to the following Euler integrals
\begin{equation}
\begin{aligned}
    \frac{n+1}{1-n}J &= (n+1)^{1-\frac{2}{n}}
    \int \zeta^{\frac{1}{n}-1}(1-\zeta)^{-\frac{1}{n}}\mathrm{d}\zeta\\
    &= \varsigma\frac{n}{(n+1)^{\frac{2}{n}-1}}
    (1-\zeta)^{1-\frac{1}{n}}\zeta^{\frac{1}{n}}\, 
    _2F_1\left(1,1,1+\tfrac{1}{n},\zeta\right)+C_2\\
    &= \varsigma'\frac{n(n+1)^{1-\frac{2}{n}}}{n-1}
    \zeta^{\frac{1}{n}}(1-\zeta)^{1-\frac{1}{n}}\, 
    _2F_1\left(1,1,2-\tfrac{1}{n},1-\zeta\right)+\widetilde{C}_2.
\end{aligned}
\end{equation}
A difference with the previous situation is that $|\zeta|$ might be greater than 1, even when we take $x=|u|^{n-1}$. Still, the above hypergeometric functions can be continued along the real axis past $\zeta=-1$ and the only potential problematic points are $\zeta=1$ and $\zeta=\infty$. The former corresponds to $f_1=f_2\iff F_2=0$,
and the latter to $f_2=0\iff F_3=0$, so they are both invariant sets. In the original variables, they become $\omega_0^2(u+\beta\dot{u})+u^n=0$ and $\beta\dot{u}+u=0$, respectively. These curves separate the phase space into regions in which one of the above hypergeometric function can be chosen. Finally, substituting for $\zeta$, $x$ and $y$, the sign factors can be determined as in the previous proof, and turn out not to depend on the region giving the stated formulae. Lastly, the integrals might acquire a complex phase due to fractional powers of 
$V = u^n+\omega_0^2(u+\beta\dot{u})$, but since this is just a multiplicative constant, its absolute value can be taken on each side of $V=0$.\qed{}

Here again, the integral itself can be written as the incomplete beta function
\begin{equation}
    \int_0^z\zeta^{\frac{1}{n}-1}(1-\zeta)^{-\frac{1}{n}}\mathrm{d}\zeta =
    B_{z}\left(\tfrac{1}{n},1-\tfrac{1}{n}\right),
\end{equation}
although it is more convenient to use the Gauss function, so that the fractional power of $V$ can be written as a common factor. In fact, this beta function turns out to be elementary, albeit with no simple general formula, as is shown in Section 5.

\begin{figure}[ht]\centering
\includegraphics[width=.45\textwidth]{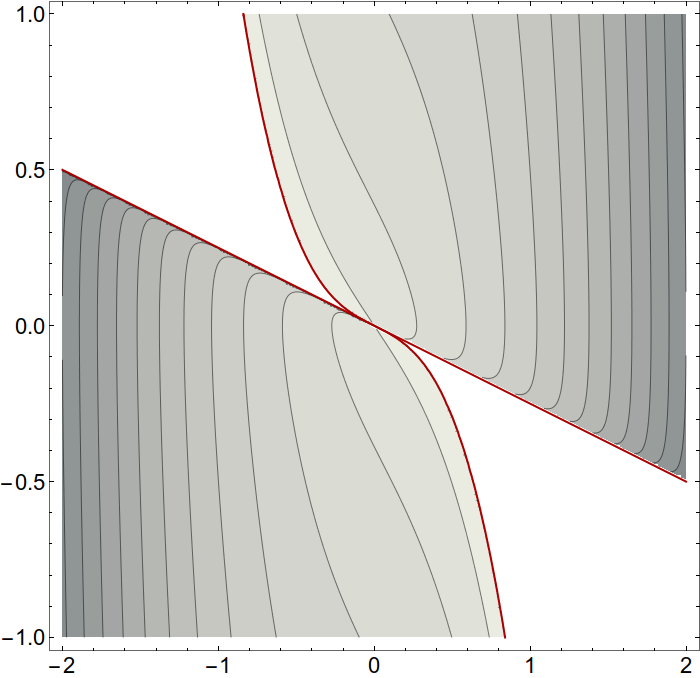}\hspace{5mm}
\includegraphics[width=.45\textwidth]{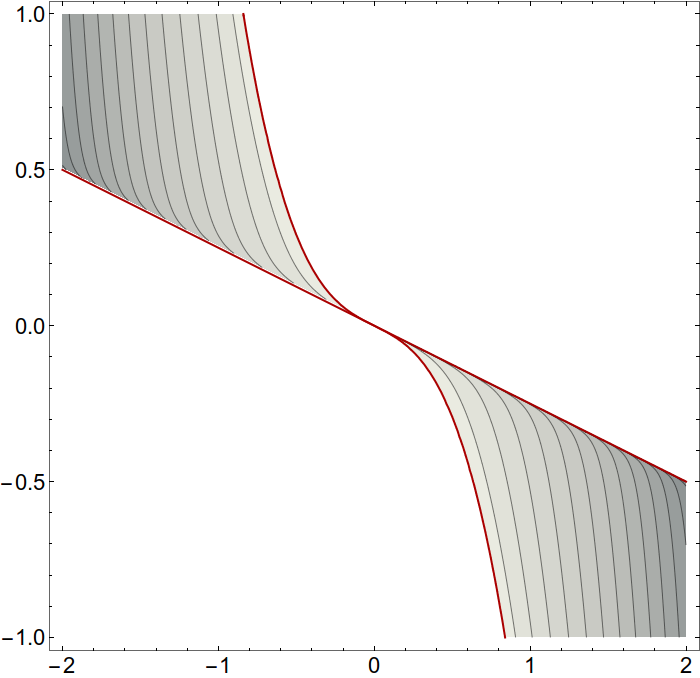}
\caption{\small The phase space of the Duffing-van der Pol oscillator with $n=3$ in the $u$ and $\dot{u}$ coordinates. The left diagram shows level sets of $I_3$ as shaded, the right one those of $I_4$, both defined in Theorem 2.}
\label{fig2}
\end{figure}

Like before, for non-integer $n$, or to keep the sign as in the harmonic case, $u^n$ can be replaced by $|u|^{n-1}u$. A plot of the level sets of $I_3$ and $I_4$ is presented in Figure \ref{fig2}. Both first integrals are valid everywhere between the red curves, and the two formulae merely reflect the fact that they are not always real-valued. As opposed to the previous system, there is no need for the regions of definition to overlap, because the red curves, being invariant sets, are impassable barriers.

\section{}
The third system presents a new challenge, because of non-polynomial Darboux elements. To wit, in the coordinates $x=u^{n-1}$, and $y=\dot{u}/u$ it reads
\begin{equation}
\begin{aligned}
\dot{x} &= (n-1)yx,\\
\dot{y} &= -\omega_0^2 - y(1+y)-x(1+\beta y)-\varphi x^2.
\end{aligned} \label{gengen_pol}
\end{equation}
If, in addition to both the previous conditions $\beta=n+1$, and
$\omega_0^2=n(n+1)^{-2}$, the new parameter satisfies
\begin{equation}
    4\varphi = \omega_0^{-2},
\end{equation}
then, there exist three Darboux pairs:
\begin{equation}
\begin{aligned}
F_1 &=x, & K_1 &= (n-1)y,\\
F_2 &=1+\beta y+ 2\varphi x, & K_2 &= (\beta^{-1}-1)-y-(\beta/2)x,\\
F_3 &= \exp\left(2\varphi F_1/F_2\right), & K_3 &= (\beta/2)x.
\end{aligned}
\end{equation}
Moreover, they generate a time-dependent first integral through
\begin{equation}
J=F_1^{\alpha_1}F_2^{\alpha_2}F_3^{\alpha_3},\quad\text{with}\quad
DJ = \frac{n(1-n)}{1+n}J,
\end{equation}
where the exponents were chosen to be
\begin{equation}
    (\alpha_1,\alpha_2,\alpha_3) =
    (1,n-1,n-1).
\end{equation}

The appearance of an exponential element along with polynomials is not surprising: these are the only two types that can arise for a two-dimensional system with a Liouvillian integral, as proved by Christopher \cite{Christopher}. It is thus natural to include exponentials, but at the same time one would wish for a criterion or procedure which, like Lemma 1, works in any dimension.

The first step in getting there is to recall that the cofactor $k_1-k_2$ had to be reexpressed as a function of $f_i$; likewise, the time-dependent integral contained $f_3$ which had to be eliminated. In other words, we are really working with Darboux elements like $f_1/f_2$, $J$ or $(f_1-f_2)^{\alpha_3}$, and they can be grouped as we please, with the aim of integration in the variable $\zeta$. This suggests the following generalization:

\begin{lmm}
Let $D=\sum_{i=1}^{d}P_i\partial_i$ be a derivation which admits at least two Darboux elements which can be combined to yield Darboux pairs $\{f_1,k_1\}$ and $\{f_2,k_2\}$ such that:\\
{\bf 1)} the element $J:=f_1 f_2$ has a constant cofactor $\alpha_0$;

\noindent{\bf 2)} the cofactor of $\zeta:=f_1/f_2$ satisfies
\begin{equation}
    k_1 - k_2 = f_2^{2\gamma}L(\zeta),\quad \mathbb{C}\ni\gamma\neq 0,
\end{equation}
for some Liouvillian $L$. Then, the dynamical system $\dot{x}=P_i(x)$ associated with the derivation $D$ has a Liouvillian first integral.

In particular, a binomial $L(\zeta) = \zeta^a(1-\zeta)^b$ leads to the Gauss function, while $L(\zeta) = \zeta^a\mathrm{e}^{b\zeta}$ to the Kummer function in the first integral.
\end{lmm}
{\it Proof.} The element $J$ considered as a function of time satisfies $J(t)=\mathrm{e}^{\alpha_0 t}J(0)$, so the integral of $J^{\gamma}$ can be formally transformed, using $\zeta=f_1/f_2$ as follows
\begin{equation}
\begin{aligned}
    \int J^{\gamma}\mathrm{d}t 
    &= \int\frac{f_1^{\gamma}f_2^{\gamma}}{(k_1-k_2)\zeta}
    \mathrm{d}\zeta
    =\int \frac{(f_1/f_2)^{\gamma}f_2^{2\gamma}}{f_2^{2\gamma}\zeta L(\zeta)}
    \mathrm{d}\zeta\\
    \frac{J^{\gamma}}{\gamma\alpha_0} - C &=\int \zeta^{\gamma-1}L(\zeta)^{-1}
    \mathrm{d}\zeta.
\end{aligned}
\end{equation}
The same considerations of complex arguments as in Lemma 1 apply, so a suitable constant $\varsigma$ of modulus 1 will have to be added once a particular region of $\zeta$ is fixed. The two special cases then give the time-independent first integral $I$ through
\begin{equation}
\begin{aligned}
    \frac{J^{\gamma}}{\varsigma\gamma\alpha_0} - I =
    \int\zeta^{\gamma-a-1}(1-\zeta)^{-b}\mathrm{d}\zeta
    &= B_{\zeta}\left(\gamma-a,1-b\right)\\
    &= \frac{\zeta^{\gamma-a}}{\gamma-a}\,
    {}_2F_1(\gamma-a,b,1+\gamma-a,\zeta),
\end{aligned}
\end{equation}
or
\begin{equation}
\begin{aligned}
    \frac{J^{\gamma}}{\varsigma\gamma\alpha_0} - I =
    \int\zeta^{\gamma-a-1}\mathrm{e}^{-b\zeta}\mathrm{d}\zeta
    &= - b^{a-\gamma}\Gamma(\gamma-a,b\zeta)\\
    &= \frac{\zeta^{\gamma-a}}{\gamma-a}\,
    {}_1F_1\left(\gamma-a,\gamma-a+1,-b\zeta\right),
\end{aligned}
\end{equation}
where $\Gamma(s,z)$ is the incomplete gamma function. \qed{}

Both the previous proofs are special cases, and they are important preliminary results showing how to combine the Darboux polynomials. In Lemma 1, denoting the ``old'' polynomials by $g_i$, we can take
$f_1 = g_1^{(\alpha_1+1)/2}g_2^{(\alpha_2-1)/2}g_3^{\alpha_3/2}$, and $f_2=g_1^{(\alpha_1-1)/2}g_2^{(\alpha_2+1)/2}g_3^{\alpha_3/2}$, so that $\zeta = f_1/f_2=g_1/g_2$, and $J=f_1 f_2 = g_1^{\alpha_1}g_2^{\alpha_2}g_3^{\alpha_3}$, while the relation between the cofactors yields
\begin{equation}
    f_2^2\zeta^{1-\alpha_1}
    \left(1-\zeta\right)^{\alpha_1+\alpha_2} =
    g_3^{\alpha_1+\alpha_2+\alpha_3} = g_2-g_1 = k_1 - k_2.
\end{equation}
when $\alpha_1+\alpha_2+\alpha_3=1$ as in the proof of Lemma 1. Similarly in the proof of Theorem 1, we could take $f_1 = F_1 F_2^{(n-3)/4}$, $f_2=F_2^{(n+1)/4}$, and then
\begin{equation}
    k_1-k_2 = K_1 - K_2 = \sqrt{\tfrac{n+1}{2}(F_2-F_1)} = \sqrt{\tfrac{n+1}{2}}f_2^{\frac{2}{n+1}}
    \sqrt{1-\zeta}.
\end{equation}
The crucial quantity is always $k_1-k_2$, which is polynomial in the dynamical variables, but not necessarily a polynomial, or even algebraic, in $f_i$.  This is ostensibly so in the present system \eqref{gengen_pol}, to which we now turn.

{\it Proof of Theorem 3.} The exponential factor suggests the choice of $\zeta=F_1/F_2$, and Lemma 2 can be used with
\begin{equation}
    f_1 = F_1 F_2^{\frac{n-2}{2}}F_3^{\frac{n-1}{2}},\quad
    f_2 = F_2^{\frac{n}{2}}F_3^{\frac{n-1}{2}}.
\end{equation}
The relation between the cofactors is
\begin{equation}
    k_1 - k_2 = K_1 - K_2 = f_2^{\frac{2}{n}}
    \exp\left(\frac{2(1-n)\varphi}{n}\zeta\right),
\end{equation}
so, by Lemma 2, the first integral is
\begin{equation}
    C = \frac{n+1}{1-n}J^{\frac{1}{n}} - \varsigma(n+1)\zeta^{\frac{1}{n}}
    {}_1F_1\left(\frac{1}{n},1+\frac{1}{n},
    \frac{2(n-1)\varphi}{n}\zeta\right).
\end{equation}
or, going back to the original variables,
\begin{equation}
    I = V^{-\frac{1}{n}}\left[\exp\left(\frac{2(n-1)\varphi u^n}{nV}\right)V+
    \varsigma (n-1)u{}_1F_1\left(\frac{1}{n},1+\frac{1}{n},
    \frac{2(n-1)\varphi u^n}{nV}\right)\right],
\end{equation}
where $V=u+\beta\dot{u}+2\varphi u^n$. The curve given by $V=0$ is invariant, so the dynamics is separated into two regions, in which the Kummer function is analytic and real. To get a single real formula it is thus enough to take the absolute value in the $V^{-1/n}$ factor. Differentiating then gives $\varsigma=1$ as in the theorem.\qed{}

\begin{figure}[ht]\centering
\includegraphics[width=.45\textwidth]{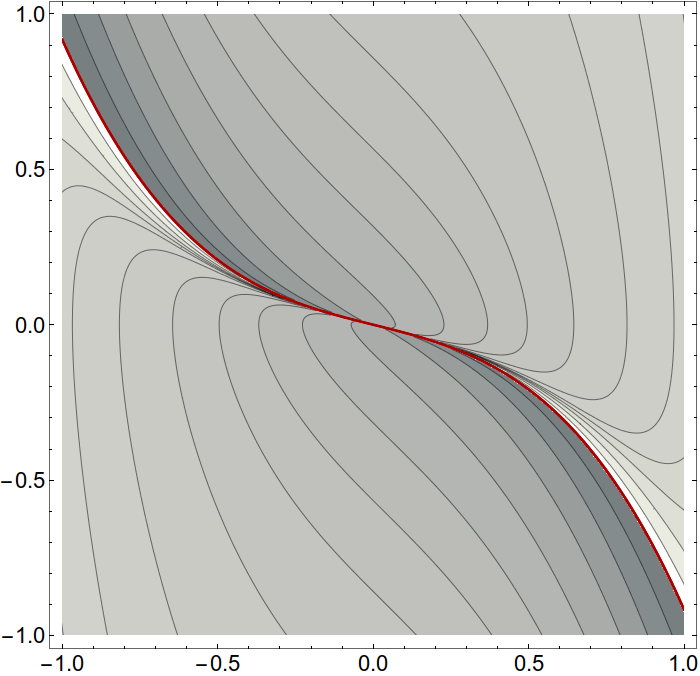}\hspace{5mm}
\includegraphics[width=.45\textwidth]{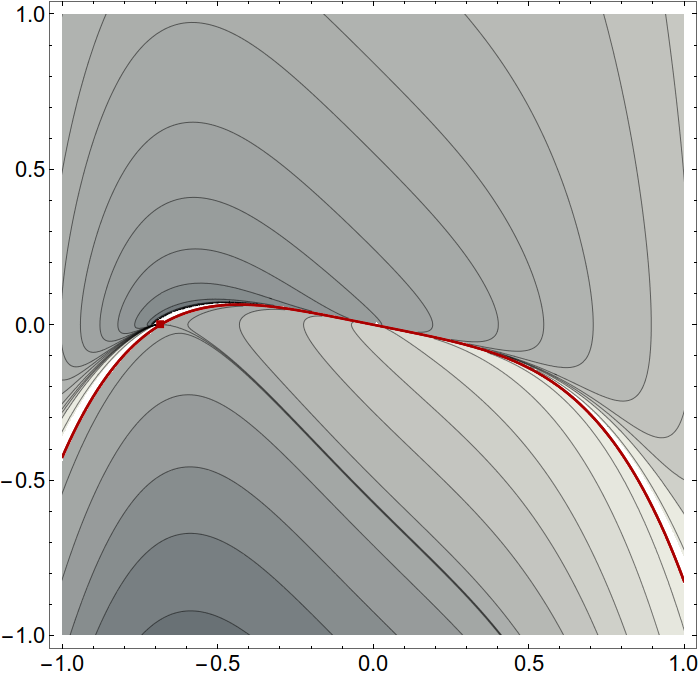}
\caption{\small The phase space of the generalized Duffing-van der Pol oscillator for $n=3$ (left) and $n=4$ (right) in the $u$ and $\dot{u}$ coordinates. All contours are given by $I_5$ of Theorem 3, the red curve is the invariant curve at the integral's singularity, and the additional critical point is indicated with a square.}
\label{fig3}
\end{figure}

Phase portraits for this system with $n=3$ and $n=4$ are shown in Figure \ref{fig3}. As before, if one wishes to consider arbitrary $n$, or just to keep a single critical point at the origin, a change of $u^n$ to $|u|^{n-1}u$ is necessary. The invariant curve corresponds to the irregular singularity of the Kummer function at infinity, so a series expansion around it is not readily available due to the Stokes phenomenon. Still, as with the exponential function, the series around zero has infinite radius of convergence.

\section{}

Having obtained the hypergeometric expressions, the natural question to ask is whether they can be reduced to elementary functions. In the case of ${}_2F_1$, the answer follows from a result of Chebyshev's \cite{Ritt}:

\begin{thm}
If $p$, $q$ and $r$ are rational numbers and $a$ and $b$ are nonzero real numbers, the indefinite integral $\int z^p(a+bz^r)^q\mathrm{d}z$ is elementary if and only if at least one of $(p+1)/r$, $q$, or $q+(p+1)/r$ is an integer.
\end{thm}
In the first system, $a=-b=r=1$ while $p=-1+1/(n+1)$ and $q=-1/2$, and the three numbers to check are $1/(n+1)$, $-1/2$, and $-1/2+1/(n+1)$ -- none of them are integers when $n>1$, so the hypergeometric function is truly transcendental in this case.

In the second system, $b=-a=r=1$ while $p=-1+1/n$ and $q=-1/n$, which means that $q+(p+1)/r=1$ regardless of $n$. However, the theorem still requires that the exponents be rational, and the specific change of variables that make the integrand rational depends on those exponents. In the central case of integer $n>1$ the change is $\zeta = 1/(1+z^n)$, and the relevant expression is
\begin{equation}
    I_z = \int \zeta^{\frac{1}{n}-1}(1-\zeta)^{-\frac{1}{n}}\mathrm{d}\zeta
    =-\int\frac{n z^{n-2}}{z^n+1}\mathrm{d}z.
\end{equation}
The evaluation of such an integral is straightforward for a specific value of $n$, e.g. $n=2$ gives $I_z=-2\text{arcsin}\left(\sqrt{1-\zeta}\right)$, but it cannot be given as a simple expression for symbolic $n$ -- other than by using the hypergeometric or beta functions.

Additionally, the transcendental functions of the first theorem can become inverses of elliptic functions. The three cases when this happens for $B_{\zeta}\left(\frac{1}{n+1},\frac{1}{2}\right)$ are given in \cite{Maier}, and correspond to $n=2,3,5$. This is best seen on the initial case, which admits a change to the Hamiltonian system \eqref{hamil}, and can be integrated for energy $E$ by
\begin{equation}
    Z = \int\frac{\sqrt{2}\,\mathrm{d}W}{\sqrt{4E-W^4}} =
    \frac{B_{\zeta}\left(\frac14,\frac12\right)}{4E^{1/4}},
\end{equation}
where, in the transformed variables, $\zeta = W^4/(4E)$. This is an inversion of the direct solution of the Hamiltonian system in terms of the Jacobian elliptic function $W=\sqrt{2}E^{1/4}\mathrm{sn}(E^{1/4}Z)$.

Finally, the Kummer function of Theorem 3 is transcendental for integer $n>1$, because it can be rewritten as the classical exponential integral 
\begin{equation}
    \Gamma\left(\tfrac{1}{n},\zeta\right) = n \int_s^{\infty}\mathrm{e}^{-s^n}\mathrm{d}s,
\end{equation}
where $\zeta = s^n$. The Gauss error function is a familiar example of the above for $n=2$.

\vspace{5mm}

\noindent{\large{\bf Acknowledgements}}

This work was supported partially by the grant No. DEC-2013/09/B/ST1/04130 of the National Science Centre of Poland, and partially by the Japan Society for the Promotion of Science, Grant-in-Aid for Scientific Research (B) (Subject No. 17H02859).

\end{document}